\documentclass[conference,a4paper]{IEEEtran}
\usepackage{amssymb,amsmath}
\usepackage{cite}
\usepackage{graphicx,subfigure}
\usepackage{psfrag}
\usepackage{url}
\usepackage[latin1]{inputenc}
\usepackage[absolute,overlay]{textpos}

\DeclareMathOperator*{\argmax}{arg\,max}
\usepackage[latin1]{inputenc}

\usepackage{algorithm}
\usepackage{algpseudocode}

\usepackage{amsthm}

\usepackage{amsmath}
\usepackage{amssymb}
\usepackage{bbm}
\usepackage{bm}
\usepackage{comment}
\usepackage[capitalize]{cleveref}

\usepackage{algorithm}
\usepackage{algpseudocode}

\begin{document}

\title{Traffic Prediction Based Fast Uplink Grant for Massive IoT}
\author{
	\IEEEauthorblockN{Mohammad Shehab$^1$, Alexander K. Hagelskj\ae r$^2$, Anders E. Kal\o r$^2$, Petar Popovski$^2$, and Hirley Alves$^1$ \\
	}	
	\IEEEauthorblockA{$^1$Centre for Wireless 	Communications (CWC), University of Oulu, Finland\\
	$^2$Department of Electronic Systems, Aalborg University, Denmark}}
\maketitle

\begin{abstract}
This paper presents a novel framework for traffic prediction of IoT devices activated by binary Markovian events. First, we consider a massive set of IoT devices whose activation events are modeled by an On-Off Markov process with known transition probabilities. Next, we exploit the temporal correlation of the traffic events and apply the forward algorithm in the context of hidden Markov models (HMM) in order to predict the activation likelihood of each IoT device. Finally, we apply the fast uplink grant scheme in order to allocate resources to the IoT devices that have the maximal likelihood for transmission. In order to evaluate the performance of the proposed scheme, we define the regret metric as the number of missed resource allocation opportunities. The proposed fast uplink scheme based on traffic prediction outperforms both conventional random access and time division duplex in terms of regret and efficiency of system usage, while it maintains its superiority over random access in terms of average age of information for massive deployments.
\end{abstract}

\section{Introduction}\label{sec:introduction}
The advent of the Internet of Things (IoT) has led to a surge in the number of devices aiming to realize the 2030 vision of data driven societies \cite{6g}. This vision is mainly dominated by a large number of machine type devices, which will be used for environment monitoring, remote surgery, autonomous objects, and yet many unforeseen applications. Several of these services require strict end-to-end QoS guarantees such as ultra-reliable and near-instant connectivity for massive machine type communication (MTC) networks \cite{MTC_6G}. Hence, an extremely low \emph{per-link} delay, in the sub-millisecond range, will be a necessary requirement in order to maintain the aggregate end-to-end latency in the order of few milliseconds.

A central element in supporting these devices is the design of protocols with efficient access procedures that are suited for spontaneous transmissions. Traditional massive access schemes such as random access (RA) and time division duplex (TDD) suffer from significant shortcomings that render them unsuitable for applications with strict requirements to both latency and reliability. For example, in conventional LTE/LTE-A RA scheme \cite{RA}, devices which have packets to transmit access the channel resources in a slotted ALOHA fashion by selecting a random slot in which they send scheduling requests to the base station. In return, the base station schedules the successfully decoded requests to available transmission slots. This approach suffers from large signalling overhead and collisions between scheduling requests that lead to longer delays and higher probability of packet drop. At the opposite side is the uncoordinated random access \cite{unc}, where the devices attempt to transmit their data in randomly selected transmission slots without relying on any scheduling procedure. This results in large number of collisions which leads to waste of power and longer delays. Meanwhile, alternative schemes such as time division duplex (TDD) and access class barring (ACB) \cite{ACB} were concluded to be inefficient in terms of delay.

The road to extreme low latency urges the invention of novel massive access schemes that depart from the shortcomings associated with the classic ones and exploit activation correlation and the prediction of traffic patterns. To illustrate traffic correlation, we consider the following example: let event 1 and event 2 correspond to the existence of fire, and someone who smokes a cigarette, respectively. Meanwhile, sensor 1 and sensor 2 are heat and smoke detectors, respectively. In case of fire, both sensors will detect the event. However, in case of smoking a cigarette, event 2 will only be detected by sensor 2. Hence, we infer that if sensor 1 is active, sensor 2 will be active with high probability but not vice versa. This scenario illustrates the ability to forecast the activation of a certain sensor when observing the activation of another one, and hence, allocating resources to the sensor which is anticipated to transmit. In this context, Fast Uplink (FU) grant was introduced in \cite{FU} to allow for resource allocation according to traffic prediction.  
\subsection{Fast Uplink Grant}

To elaborate on the FU grant, consider the availability of $L$ transmission slots and $K$ IoT devices where $K\gg L$ and each device is triggered to generate packets at different times depending on different events. When a device generates a packet, it requires a transmission slot to transmit this packet. In the FU grant scheme, the common aggregator allocates available transmission slots to IoT devices based on a traffic prediction scheme, where the learning procedure exploits the correlation of traffic patterns on the temporal and event dimensions. The advantages of the FU grant scheme are:
\begin{itemize}
\item Reduced power consumption of the devices and decreased latency since there are no scheduling requests and no collisions;
\item No signalling overhead between the devices since learning occurs only at the side of the aggregator;
\item Permits usage of the uplink grant signal in order to partially or fully enable channel estimation at the IoT devices before uplink (CSIT);
\end{itemize}
Note that the channel estimation is a proposed advantage of applying the FU grant scheme that could be investigated for possible future implementation.

\begin{figure}[!t] % [!t] or [!b] or [!h] % force fitting, force top, force bottom, force text fitting
		\centering 
		\includegraphics[trim=1cm 15cm 1.5cm 1cm, clip=true,width=1\linewidth]{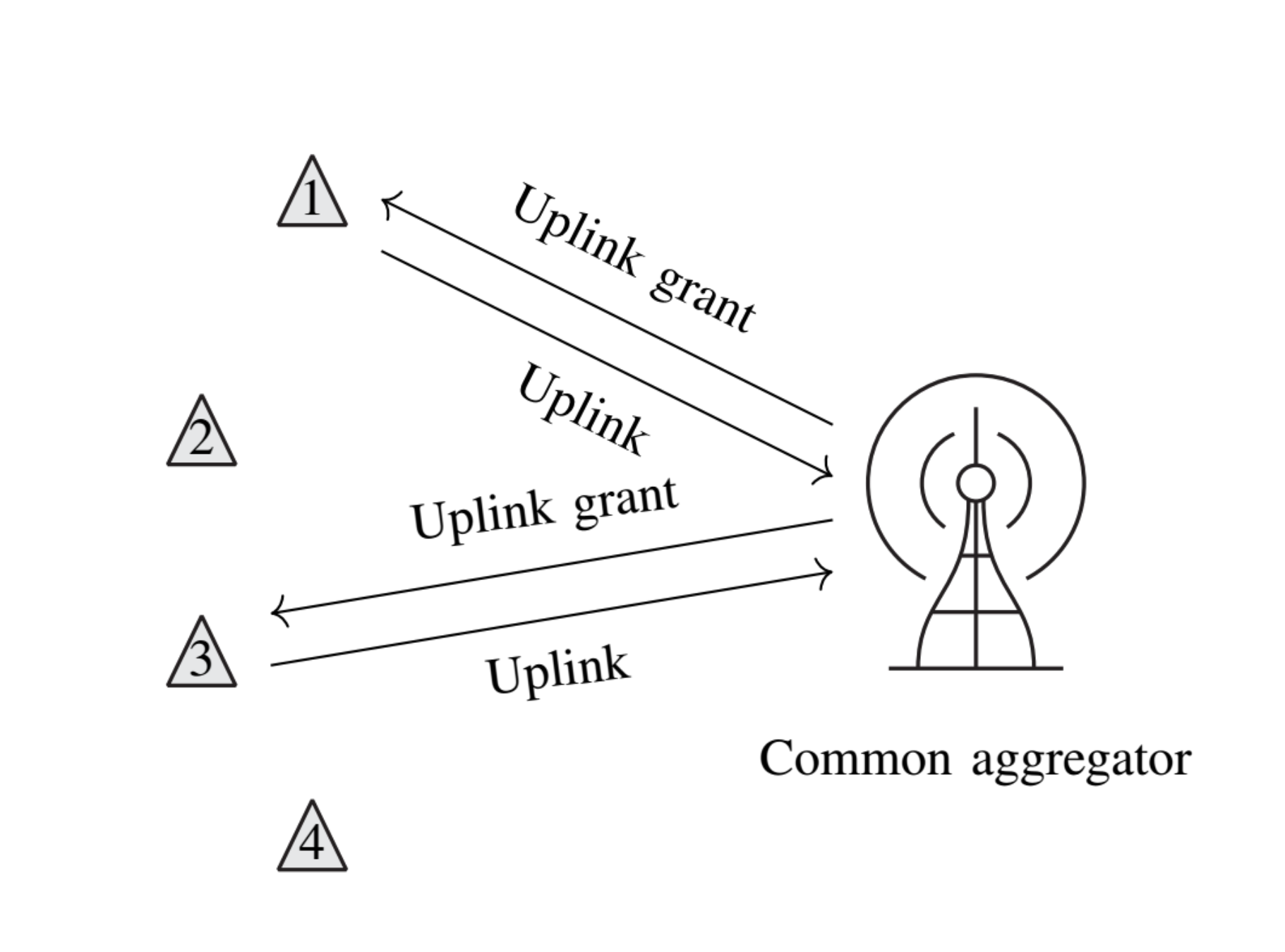}
		\vspace{-2mm}
		\caption{Fast Uplink scenario. The common aggregator grants access to devices 1 and 3, while 2 and 4 remain inactive.}
		\label{FU}
		\vspace{-2mm}
	\end{figure}

The proposed FU grant model is depicted in Fig. \ref{FU} for $k=4$ devices and $L=2$ slots at a certain time instant, where only 2 out of 4 devices are active. The FU approach suggests that the aggregator is able to predict the likelihood of the traffic pattern of the 4 devices and grants the 2 available transmission slots to the 2 devices which are more likely to transmit. In \cite{Samad_MAB}, Samad et. al proposed a multi-armed bandit framework to perform fast uplink grant for IoT devices. However, the authors did not exploit the traffic correlation on the event-temporal dimensions. Meanwhile, the authors of \cite{Anders_RA} exploited the correlated activity of devices to develop heuristic RA protocols. Moreover, works such as \cite{CM1,CM2} characterized device activations according to coupled Markovian traffic models but without addressing the resource allocation problem based on these traffic models. 

Herein, we assume discrete  events that induce the activity of IoT devices and such events are modeled based on an On-Off Markov arrival process. Then, we develop an efficient traffic prediction scheme which exploits the traffic correlation on the event and temporal dimensions in order to predict the likelihood of transmission of each device and enable the FU grant procedure. 

	\begin{figure}[!t] % [!t] or [!b] or [!h] % force fitting, force top, force bottom, force text fitting
		\centering 
		\includegraphics[trim=3cm 13.5cm 3cm 1cm, clip=true,width=1\linewidth]{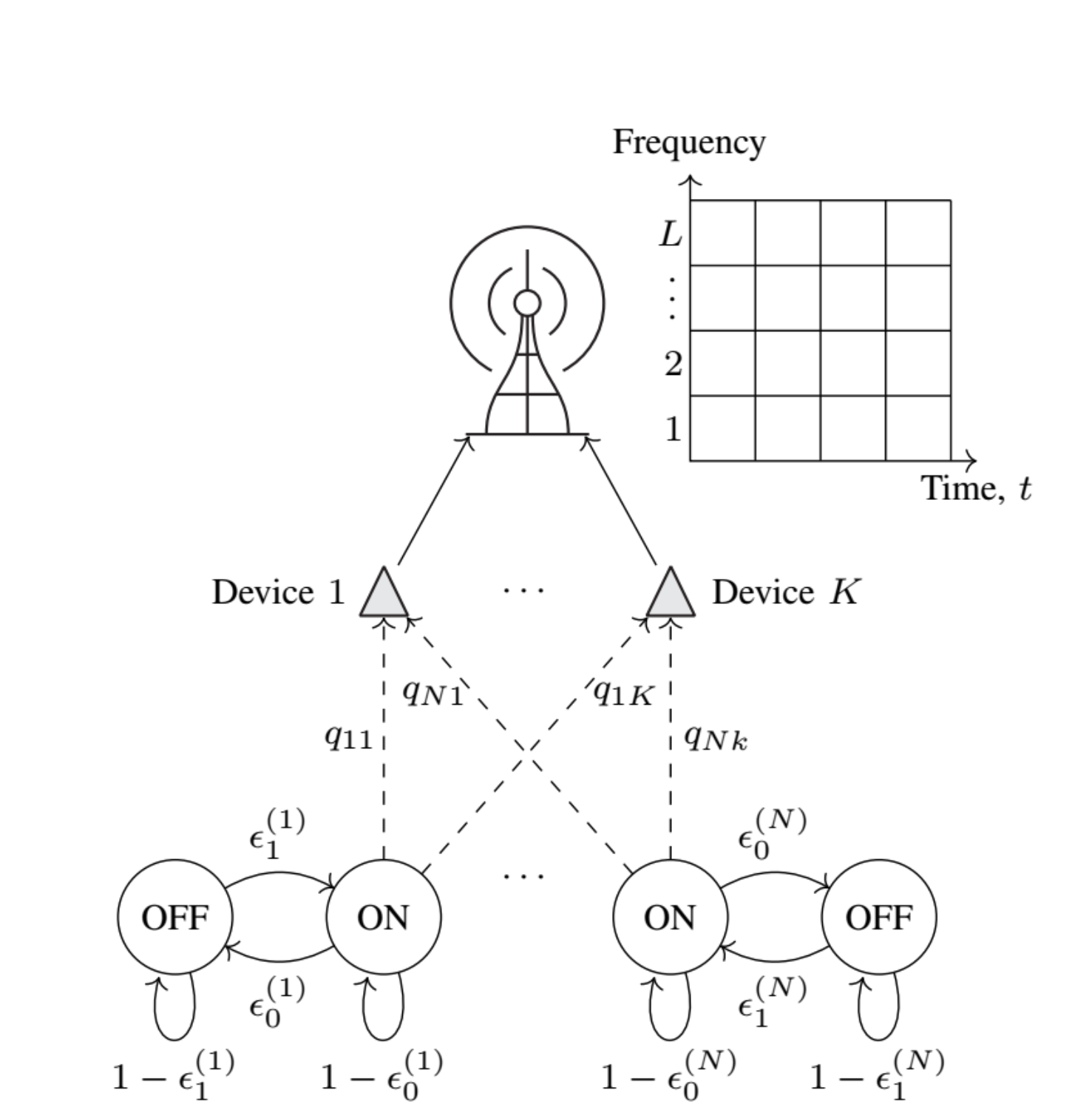}
		\vspace{-2mm}
		\caption{The considered activation model, in which $N$ On-Off processes control the activation of $K$ devices. If process $n$ is in the On-state it activates device $k$ with probability $q_{nk}$.}
		\label{fig:scenario}
		\vspace{-2mm}
	\end{figure}

\subsection{Outline}
The rest of the paper is organized as follows: in Section~\ref {sec:sysmodel}, we introduce the system model for the IoT device stimulated by Markovian events. Next, Section~\ref{sec:sysanalysis} proposes the forward algorithm which is applied in order to predict the traffic pattern of IoT devices. It also introduces performance metrics which will be used to evaluate the performance of the proposed fast uplink scheme. After that, Section~\ref{results} shows and discusses different results for the performance evaluation. Finally, Section~\ref{conclusions} concludes the paper and discusses future research directions. 

\section{System model}\label{sec:sysmodel}
We consider the uplink model depicted in \cref{fig:scenario}, with $K$ IoT devices and a single aggregator. The radio link is divided into time slots, and in each time slot the base station can schedule up to $L$ devices for transmission in $L$ transmission slots, as in LTE fast uplink. The scheduled devices are allotted dedicated resources, and transmit only if they are active, i.e. if they have data to transmit. 
If an inactive device is scheduled for transmission, the uplink resource is wasted.

We denote the activation of device $k$ in discrete time slots $t=1,2, \dots$ by the random variable $A_t^{(k)}$, which is equal to one if the device is active and zero otherwise. The IoT devices activation vector at time $t$ is given by $\textbf{A}_t = \left\lbrace A_t^{(1)}, ..., A_t^{(K)} \right\rbrace$. 

The activation pattern of the devices is controlled by $N$ independent two-state Markov processes. Each Markov process is characterized by On and Off states, where the state at time $t$, $\mathcal{S}_t^{(n)} \in \ \left\lbrace 1,0\right\rbrace $, is governed by temporal transition probabilities $\epsilon_1^{(n)}$, $\epsilon_0^{(n)}$ as shown in Fig. \ref{fig:scenario}, where
\begin{equation}
    \Pr\left( \mathcal{S}_{t+1}^{(n)}=0 \middle| \mathcal{S}_t^{(n)}=1\right) =\epsilon_0^{(n)}, 
\end{equation}
\begin{equation}
    \Pr\left( \mathcal{S}_{t+1}^{(n)}=1 \middle| \mathcal{S}_t^{(n)}=0\right) =\epsilon_1^{(n)}, 
\end{equation}
\begin{equation}
    \Pr\left( \mathcal{S}_{t+1}^{(n)}=0 \middle| \mathcal{S}_t^{(n)}=0\right) =1-\epsilon_1^{(n)}, 
\end{equation}
\begin{equation}
    \Pr\left( \mathcal{S}_{t+1}^{(n)}=1 \middle| \mathcal{S}_t^{(n)}=1\right) =1-\epsilon_0^{(n)}. 
\end{equation}
Moreover, we define the state vector at time $t$ as $\mathcal{\textbf{S}}_t=\left\lbrace \mathcal{S}_t^{(1)},...,\mathcal{S}_t^{(N)}\right\rbrace $.

The Markov processes that are in the On state, i.e. $\mathcal{S}_t^{(n)}=1$, may activate certain IoT devices. More specifically, the probability that Markov process $n$ activates device $k$ in the On state is $q_{nk}$.

\section{System Analysis}\label{sec:sysanalysis} \vspace{1mm}
In this section, we analyze the device temporal activation probabilities and exploit them develop the traffic prediction based FU scheme. Next, we introduce the notion of regret, system usage, and age of information (AoI) as performance metrics. These metrics are useful in the evaluation of the performance of the proposed FU scheme.
\subsection{Device Activation Probabilities}
A device is active if any of the $N$ Markov processes activates the device, so that the probability that device $k$ is active at time $t$ is 
\begin{align}
   \Pr\left( A_{t}^{(k)}=1 \middle| \mathcal{\textbf{S}}_t\right)&=1-\bigcap_{n=1}^N \Pr\left( A_{t}^{(k)}=0 \middle | \mathcal{S}_t^{(n)}  \right) \\
   &=1-\prod_{n=1}^N (1-q_{nk})^{\mathcal{S}_t^{(n)} },
\end{align}
where we resorted to the fact that the activation is conditionally independent given the state vector $\mathbf{S}_t$. 

Moreover, the probability of IoT device $k$ to be active at time $t+1$ given the state vector at time $t$ is as 
\begin{align} %\label{activation prediction}
   \Pr\left( A_{t+1}^{(k)}=1 \middle| \mathcal{\textbf{S}}_t\right)&=1-\bigcap_{n=1}^N \Pr\left( A_{t+1}^{(k)}=0 \middle | \mathcal{S}_t^{(n)}  \right) \\
   &=1-\prod_{n=1}^N h(n),  \label{activation prediction}
\end{align}
where
\begin{equation}
   h(n)=
      \begin{cases}
        1-\epsilon_1^{(n)}+\epsilon_1^{(n)}(1-q_{nk}), & \quad \mathcal{S}_t^{(n)}=0\\
       \epsilon_0^{(n)}+(1-\epsilon_0^{(n)})(1-q_{nk}), & \quad  \mathcal{S}_t^{(n)}=1.
     \end{cases}
\end{equation}
\vspace{2mm}
\subsection{Traffic Pattern Prediction}
The states of the Markov processes are generally unknown to the base station, which has to continuously estimate them based on the device activation. To this end, we exploit that the activation observed by the base station can be described by an $N$-Hidden Markov Model (HMM) \cite{HMM}, where the total number of possible states at a certain time slot is $2^N$. In particular, we apply the forward algorithm to determine the probability of the process being in a state, given a history of observations, and use the resulting state distribution to predict future device activation.

The forward algorithm computes the joint probability $p(\mathbf{S}_t, \mathbf{A}_t)$ efficiently by formulating the problem recursively \cite{HMM2}. In the notation of this paper, the forward algorithm can be formulated as
\begin{equation} \label{alpha_equation}
p(\mathcal{\textbf{S}}_t,\textbf{A}_{1:t}) = p\left( \textbf{A}_{t} \middle| \mathcal{\textbf{S}}_t \right) \sum_{\textbf{S}_{t-1}} p\left( \textbf{S}_t \middle|  \textbf{S}_{t-1} \right) p(\mathbf{S}_{t-1}, \mathbf{A}_{1:t-1}).
\end{equation}
This joint probability can be used for predicting the most likely state, given a series of observations as
\begin{equation}\label{eq:most_likely_S}
\textbf{S}_t^*=\argmax_{\textbf{S}_t} ~ p(\mathcal{\textbf{S}}_t,\textbf{A}_{1:t}) , 
\end{equation}
or, similarly, for predicting the most likely set of active devices in the next time step
\begin{equation}\label{eq:most_likely_A}
\textbf{A}_{t+1}^*=\argmax_{\textbf{A}_{t+1}}  \prod_{k=1}^K \Pr\left( A_{t+1}^{(k)}=b_k \middle| \textbf{S}_t^* \right).
\end{equation}
Here $\textbf{A}_{t+1}^*$ is the maximum likelihood estimate of the set of active IoT devices at $t+1$ with $b_k \in \ \left\lbrace 1,0\right\rbrace$. Equation \eqref{eq:most_likely_A} can be iteratively used to predict the state of events at $t+2$ using \eqref{eq:most_likely_S} after partial correction of the activation pattern by setting the activation indicator $A_{t+1}^{(k)^*}$ to zero (limited information) for devices that were granted transmission but did not transmit.

When performing uplink grant allocation, we have $L$ slots to use at each time $t$. While the activation pattern most likely to be observed is given by \eqref{eq:most_likely_A}, we are interested in selecting the $L$ devices which are most likely to be active. Since \eqref{eq:most_likely_A} only evaluates the probability of a full pattern and does not consider the activation probability of individual devices, it becomes difficult to determine which devices in $\textbf{A}_{t+1}^*$ are most likely to be active.
To find these devices we instead assume that the system is in the most likely state, as found in \eqref{eq:most_likely_S}, and use that assumption to calculate probability that an device is active in the next time slot using \eqref{activation prediction}.
The devices are sorted by their probability of activation, and the $L$ devices most likely to be active are scheduled in the next slot.

\iffalse
\begin{equation}
    P_{\text{On}}^{(n)} = \Pr \left (\mathcal{S}_{t+1}^{(n)} = 1 \middle | \mathcal{S}_{t}^{(n)} = 1 \right ).
\end{equation}
Finally, we find the likelihood of each device activation as follows:
\begin{align}
    P_{\text{device}}^{(k)} = P_{\text{On}}^{(1)} \cdot q_{1,k} & \bigcup\\
    P_{\text{On}}^{(2)} \cdot q_{2,k} & \bigcup\\
    \ldots & \bigcup \\
    P_{\text{On}}^{(n)} \cdot q_{n,k} & .
\end{align}
\fi
 
Next, we define some performance metrics that could be useful to evaluate the performance of the proposed FU grant scheme with traffic prediction.

\subsection{Regret}
The performance of a FU scheme can be quantified using the notion of \emph{regret} \cite{FU}. Here, regret can be defined as the missed opportunity that occurs when a slot is allocated to an inactive device, while there is another device that has data to transmit but receives no grant. We consider the problem of predicting and subsequently scheduling the active devices in each time slot. To this end, regret is  cumulative number of wasted resources that could have been allocated to active unserved devices.  Hence, the regret can be considered as an indicator of the efficiency of the system usage. In order to capture the scheduling decisions at time $t$, consider the uplink grant vector $\mathbf{G}_t=\left\lbrace g_t^{(1)}, \dots, g_t^{(K)}\right \rbrace$, where $g_t^{(k)}$ is equal to 1 if a slot is allocated to device $k$ and 0 if no slot is allocated to device $k$. In this case, the number of wrong allocations is defined as the number of slots that are allocated to devices, that have no data to transmit which is
\begin{equation}
    \omega_t=\sum_{k=1}^K \left[ g_t^{(k)}-A_t^{(k)}\right]^+,
\end{equation}
where $[x]^+=\max(0,x)$.
Moreover, the number of missed allocations is defined as the number of devices which have data to transmit but do not receive a grant. That is 
\begin{equation}
    \gamma_t=\sum_{k=1}^K \left[ A_t^{(k)}-g_t^{(k)}\right]^+.
\end{equation}
Hence, we define the regret function at time $t$ as \vspace{0.5mm}
\begin{equation}
    R(t) = \min \left \lbrace \omega_t,\gamma_t \right \rbrace.
\end{equation}
Then the overall goal is to  minimize the long-term $R(t)$.

To obtain some intuition about the regret function, we consider the regret in three scenarios. In the first, suppose there are $M>L$ active devices, and that all $L$ uplink grants are given a subset of the active devices. In this case $\omega_t=0$ and $\gamma_t=0$, and so the regret $R(t)=0$, which reflects the fact that the number of unserved devices is minimized. In the opposite extreme case, $0$ devices are active, and the $L$ grants are given to inactive devices. In this case $\omega_t=L$ and $\gamma_t=0$, and thus $R(t)=0$ also in this case, which again reflects the fact that the number of unserved devices is minimized. Now, suppose that $M\le 2L$ devices are active, and that half of the active devices are given grants, while the remaining $L-M/2$ grants are given to inactive devices. We then have $\omega_t=L-M/2$ and $\gamma_t=M/2$, so $R(t)=\min(L-M/2,M/2)$, which is the number of unserved devices that could have been served by proper allocation of the grants.
\vspace{1mm}
\subsection{System usage} \vspace{1mm}
In order to quantify the efficiency of the proposed resource allocation scheme, we propose the system usage metric. The average system usage $\eta_t$ at time $t$  is simply the time averaged ratio between the number of transmission slots that are being successfully used by a transmitting device and the total number of slots $L$. That is 
\begin{equation}
    \eta_t=\frac{1}{tL}\sum_{\tau=0}^t L-\omega_\tau.
\end{equation}
The average system usage indicates the percentage of resources successfully allocated and used for transmission by IoT devices.
\vspace{1mm}
\subsection{Age of Information} \vspace{1mm}
The age of information (AoI) \cite{AoI,bedewy} of device $k$ is defined as the time elapsed since device $k$ successfully transmitted a packet, which is the last time device $k$ was activated and received transmission grant. That is
\begin{equation}
    a^{(k)}=t-t_k,
\end{equation}
where $t_k<t$ is the last time slot before $t$ when $A_{t_k}^{(k)}=g_{t_k}^{(k)}=1$. As we infer from the definition, the AoI here is a discrete process whose value is non-negative integer. Hence, the peak age per device can be defined as $\max_k \lbrace a^{(k)} \rbrace$, while the average age per device at a certain time is given by 
\begin{equation}
    \Bar{a}=\frac{1}{K}\sum_{k=1}^K a^{(k)}.
\end{equation}
The importance of the AoI metric as a performance measure in our scenario is that it measures the freshness of the data packets received from each device. That is, if a device is rarely scheduled for transmission, the data stored at the common aggregator from this sensor will be outdated as the device's age becomes too high. This could be considered as a measure of fairness where devices that are fairly scheduled from time to time would have a relatively low average age.
\vspace{0.8mm}
\section{Peformance evaluation}\label{results} \vspace{0.6mm}
Consider a setup of $N=10$ Markovian events monitored by $K=50$ sensors competing for $L=10$ frequency slots at each time slot. The temporal state transition probabilities are $\epsilon_0^{(n)}$ and $\epsilon_1^{(n)}$ are uniformly distributed on the interval $[0,0.5]$, where low values of $\epsilon$ correspond to events which stay in one state for longer times and hence cause more bursty traffic. Meanwhile the activation probabilities $q_{nk}\in[0,1]$. \vspace{0.5mm}

\begin{figure}[!t] % [!t] or [!b] or [!h] % force fitting, force top, force bottom, force text fitting
		\centering 
		\includegraphics[trim=0cm 4.5cm 1.5cm 5.3cm, clip=true,width=1\linewidth]{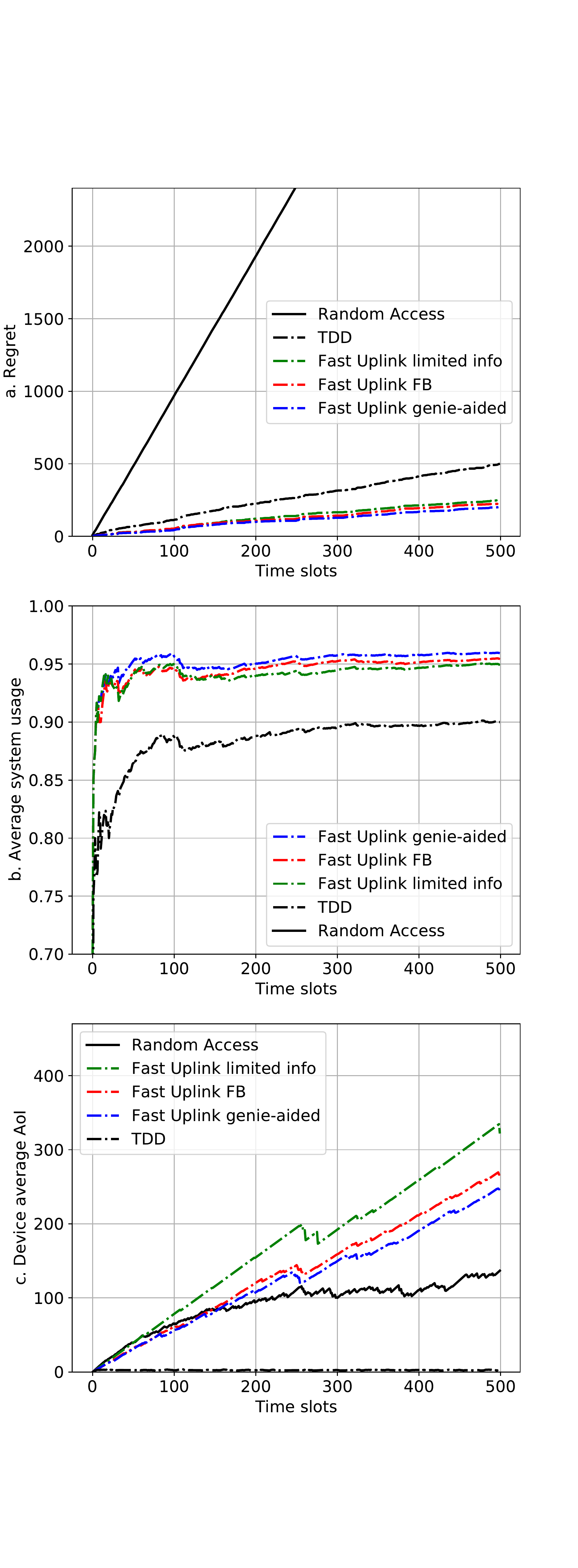}
		\vspace{-0mm}
		\caption{Performance evaluation for 10 events observed by 50 sensors competing for 10 transmission slots.}
		\label{performance1}
		\vspace{-0mm}
	\end{figure}

\begin{figure}[!t] % [!t] or [!b] or [!h] % force fitting, force top, force bottom, force text fitting
		\centering 
		\includegraphics[trim=0cm 4.5cm 1.5cm 5.3cm, clip=true,width=1\linewidth]{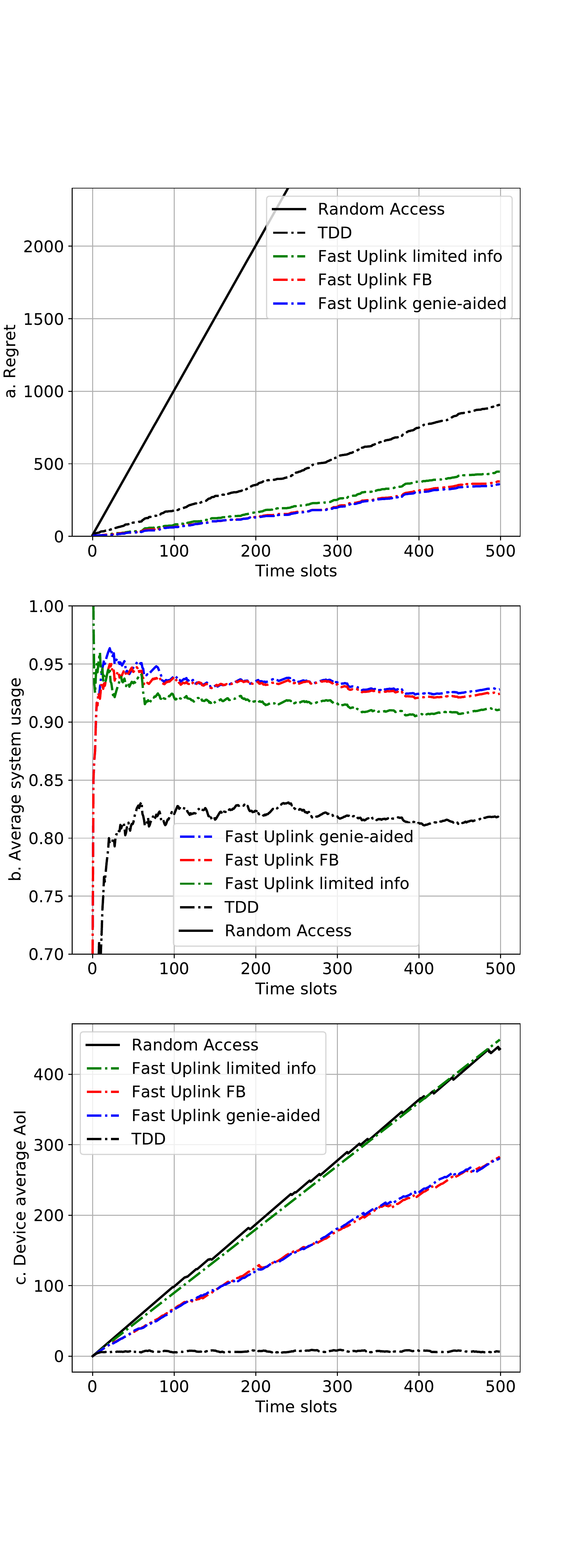}
		\vspace{-0mm}
		\caption{Performance evaluation for 10 events observed by 100 sensors competing for 10 transmission slots.}
		\label{performance2}
		\vspace{-0mm}
	\end{figure}

Fig. \ref{performance1} demonstrates different performance metrics when applying RA, TDD (round-robin), FU with limited information, FU with feedback (FB), and the genie-aided FU. Here, FU with limited information corresponds to the case in which the base station observes the activation only of the scheduled sensors, while FB allows the base station to observe, through a feedback signal, the activation also of devices that were not scheduled. Finally, the genie-aided FU refers to the the case in which he states of the sources is known to the base station. In Fig. \ref{performance1}a, we evaluate the regret function, where the FU scheme significantly outperform both RA and TDD. Specifically, when applying the proposed FU scheme, the regret function is reduced to half the regret in case of TDD and 15 times less than the regret of RA. Moreover, the FU scheme with limited info, where there is no feedback about the active devices or events, renders a very close regret to the one obtained by the genie-aided model which assumes perfect knowledge of events. The performance is also close to that of the feedback based FU which assumes knowledge of all active devices at time slot $t$ to predict the transmission likelihood at $t+1$. The same behaviour occurs for the case of $K=100$ sensors in Fig. \ref{performance2}a.

\vspace{0.5mm}

Fig. 3b depicts the average system usage for each resource allocation scheme. It is observed that the system has nearly a 0.96 system usage for the proposed FU schemes which means that an average of $96\%$ of the resource blocks are successfully allocated to transmitting devices. Thus, the proposed FU schemes are more efficient than TDD which exploits only $89\%$ of the resources. Meanwhile, RA is the worst performer due to the high number of collisions and its plot is omitted as it corresponds to nearly zero system usage. Fig. 3c shows the average age per device, where the proposed FU schemes has relatively higher ages when compared to RA and TDD. Note that the age plots capture the saw-tooth behaviour of individual devices as their ages drop down whenever they are scheduled and transmit.

Fig. \ref{performance2} demonstrates the same performance metrics for a more massive system with $K=100$ sensors. Again, it is obvious that the proposed FU schemes significantly outperforms both RA and TDD in terms of Regret and system usage. In this case, FU successfully allocates more than $92\%$ of the resource blocks to transmitting devices compared to only $81\%$. This highlights the efficiency of the proposed FU scheme in massive networks. Moreover, unlike the case of less-dense network, the proposed FU schemes performs better in terms of average age per device when compared to RA, while TDD still renders the least average age. \vspace{0.7mm}

\section{Conclusions and discussion}\label{conclusions} \vspace{0.9mm}
Efficient traffic prediction could be applied to preemptively allocate resources to IoT device via resource allocation schemes such as fast uplink (FU) grant. In this paper, we propose a traffic prediction scheme that efficiently predicts the transmission likelihood of sensors stimulated by Markovian events in massive IoT scenarios. Accordingly resource blocks are allocated to sensors that are most likely to be active at each time slot via FU grant. The proposed scheme significantly outperforms both RA and TDD in terms of regret and system usage, and this performance superiority is more obvious in more dense networks. 

For more dense networks, the average age per device of FU is lower than RA and thus, it also outperforms RA age-wise. However, for less dense networks, the average age per device for the proposed FU scheme is higher than both RA and TDD. The proposed FU grant scheme renders relatively high average age since sensors with very low transmission likelihood are rarely scheduled. This fairness issue is an open problem for future research where the target could be minimizing the regret function subject to peak age per device constraint. Finally, the benefits of efficient traffic prediction in such scenarios could be extended to the efficient allocation of other types of resources such as in optimum UAV (unmanned aerial vehicle) positioning and network design as envisioned by \cite{U3}.
%
%\vspace*{-1mm}
\section*{Acknowledgments}
This work is partially supported by Academy of Finland 6Genesis Flagship (Grant no. 318927), Aka Project EE-IoT (Grant no. 319008). This work has been in part supported by the European Research Council (ERC) under the European Union Horizon 2020 research and innovation program (ERC Consolidator Grant Nr. 648382 WILLOW) and Danish Council for Independent Research (Grant Nr. 8022-00284B SEMIOTIC).

\appendices 
%

%\bibliographystyle{IEEEtran}
%\bibliography{IEEEabrv,references}

\begin{thebibliography}{10}
\providecommand{\url}[1]{#1}
\csname url@samestyle\endcsname
\providecommand{\newblock}{\relax}
\providecommand{\bibinfo}[2]{#2}
\providecommand{\BIBentrySTDinterwordspacing}{\spaceskip=0pt\relax}
\providecommand{\BIBentryALTinterwordstretchfactor}{4}
\providecommand{\BIBentryALTinterwordspacing}{\spaceskip=\fontdimen2\font plus
\BIBentryALTinterwordstretchfactor\fontdimen3\font minus
  \fontdimen4\font\relax}
\providecommand{\BIBforeignlanguage}[2]{{%
\expandafter\ifx\csname l@#1\endcsname\relax
\typeout{** WARNING: IEEEtran.bst: No hyphenation pattern has been}%
\typeout{** loaded for the language `#1'. Using the pattern for}%
\typeout{** the default language instead.}%
\else
\language=\csname l@#1\endcsname
\fi
#2}}
\providecommand{\BIBdecl}{\relax}
\BIBdecl

\bibitem{6g}
M.~Latva-aho and K.~Leppanen, ``{Key Drivers and Research Challenges for 6G
  Ubiquitous Wireless Intelligence},'' \emph{6G Flagship, University of Oulu,
  Finland}, Sep 2019.

\bibitem{MTC_6G}
\BIBentryALTinterwordspacing
N.~H. Mahmood, H.~Alves, O.~L.~A. L{\'{o}}pez, M.~Shehab, D.~P.~M. Osorio, and
  M.~Latva{-}aho, ``{Six Key Enablers for Machine Type Communication in 6G},''
  \emph{CoRR}, vol. abs/1903.05406, 2019. [Online]. Available:
  \url{http://arxiv.org/abs/1903.05406}
\BIBentrySTDinterwordspacing

\bibitem{RA}
A.~{Laya}, L.~{Alonso}, and J.~{Alonso-Zarate}, ``{Is the Random Access Channel
  of LTE and LTE-A Suitable for M2M Communications? A Survey of
  Alternatives},'' \emph{IEEE Communications Surveys Tutorials}, vol.~16,
  no.~1, pp. 4--16, First 2014.

\bibitem{unc}
D.~{Zucchetto} and A.~{Zanella}, ``Uncoordinated access schemes for the iot:
  Approaches, regulations, and performance,'' \emph{IEEE Communications
  Magazine}, vol.~55, no.~9, pp. 48--54, Sep. 2017.

\bibitem{ACB}
3GPP, ``{Service accessibility},'' {3rd Generation Partnership Project (3GPP)},
  Technical Specification (TS) 22.011, 12 2018, version 16.4.0.

\bibitem{FU}
S.~{Ali}, N.~{Rajatheva}, and W.~{Saad}, ``Fast uplink grant for machine type
  communications: Challenges and opportunities,'' \emph{IEEE Communications
  Magazine}, vol.~57, no.~3, pp. 97--103, March 2019.

\bibitem{Samad_MAB}
S.~{Ali}, A.~{Ferdowsi}, W.~{Saad}, and N.~{Rajatheva}, ``Sleeping multi-armed
  bandits for fast uplink grant allocation in machine type communications,'' in
  \emph{2018 IEEE Globecom Workshops (GC Wkshps)}, Dec 2018, pp. 1--6.

\bibitem{Anders_RA}
A.~E. {Kal{\o}r}, O.~A. {Hanna}, and P.~{Popovski}, ``{Random Access Schemes in
  Wireless Systems with Correlated User Activity},'' in \emph{2018 IEEE 19th
  International Workshop on Signal Processing Advances in Wireless
  Communications (SPAWC)}, June 2018, pp. 1--5.

\bibitem{CM1}
M.~{Laner}, P.~{Svoboda}, N.~{Nikaein}, and M.~{Rupp}, ``{Traffic Models for
  Machine Type Communications},'' in \emph{ISWCS 2013; The Tenth International
  Symposium on Wireless Communication Systems}, Aug 2013, pp. 1--5.

\bibitem{CM2}
E.~{Grigoreva}, M.~{Laurer}, M.~{Vilgelm}, T.~{Gehrsitz}, and W.~{Kellerer},
  ``{Coupled Markovian Arrival Process for Automotive Machine Type
  Communication traffic modeling},'' in \emph{2017 IEEE International
  Conference on Communications (ICC)}, May 2017, pp. 1--6.

\bibitem{HMM}
O.~Capp, E.~Moulines, and T.~Ryden, \emph{Inference in Hidden Markov
  Models}.\hskip 1em plus 0.5em minus 0.4em\relax Springer Publishing Company,
  Incorporated, 2010.

\bibitem{HMM2}
L.~R. {Rabiner}, ``{A tutorial on hidden Markov models and selected
  applications in speech recognition},'' \emph{Proceedings of the IEEE},
  vol.~77, no.~2, pp. 257--286, Feb 1989.

\bibitem{AoI}
A.~Kosta, N.~Pappas, and V.~Angelakis, ``Age of information: A new concept,
  metric, and tool,'' \emph{Foundations and Trends in Networking, Now
  Publishers, Inc.}, 2017.

\bibitem{bedewy}
A.~M. {Bedewy}, Y.~{Sun}, and N.~B. {Shroff}, ``Minimizing the age of
  information through queues,'' \emph{IEEE Transactions on Information Theory},
  vol.~65, no.~8, pp. 5215--5232, 2019.

\bibitem{U3}
M.~{Mozaffari}, W.~{Saad}, M.~{Bennis}, and M.~{Debbah}, ``{Mobile Unmanned
  Aerial Vehicles (UAVs) for Energy-Efficient Internet of Things
  Communications},'' \emph{IEEE Transactions on Wireless Communications},
  vol.~16, no.~11, pp. 7574--7589, Nov 2017.

\end{thebibliography}


%
\end{document}